\documentclass[preprint,12pt]{elsarticle}
\usepackage[utf8]{inputenc}
\usepackage{amssymb}
\newcounter{bla}

\journal{Computer Physics Communications}

\begin{document}

\begin{frontmatter}



\title{
\begin{flushright}
{\normalsize  
arXiv:1807\\
LU-TP 18-20\\
MCNET-18-13\\
}
\end{flushright}
PISTA: Posterior Ion STAcking}


\author[lund]{Johannes Bellm\corref{author}}
\author[lund,bohr]{Christian Bierlich\corref{author}}

\cortext[author] {Corresponding author.\\\textit{E-mail address:} johannes.bellm@thep.lu.se, christian.bierlich@thep.lu.se}
\address[lund]{Dept. of Astronomy and Theoretical Physics, S{\" o}lvegatan 14A, S-223 62 Lund, Sweden}
\address[bohr]{Niels Bohr Institute, Blegdamsvej 17, 2100 Copenhagen, Denmark}

\begin{abstract}
We present a program allowing for the construction of a heavy ion Monte Carlo
event using a general purpose proton--proton event generator of the users'
choice to deliver sub--collisions. In this heavy ion event, one or more jets can
be added, again using a jet calculation of the users' choice, allowing for more
realistic simulation studies of jets in a heavy ion background.
\end{abstract}


\end{frontmatter}
% \linenumbers

% Computer program descriptions should contain the following PROGRAM SUMMARY.

{\bf PROGRAM SUMMARY}

\begin{small}
\noindent
{\em Program Title: PISTA: Posterior Ion STAcking}                  \\
{\em Licensing provisions: GPLv3}                                   \\
{\em Programming language: Python}                                   \\
{\em Supplementary material:}                                 \\
  % Fill in if necessary, otherwise leave out.
{\em Journal reference of previous version:}                  \\
  %Only required for a New Version summary, otherwise leave out.
{\em Does the new version supersede the previous version?:}   \\
  %Only required for a New Version summary, otherwise leave out.
{\em Reasons for the new version:}\\
  %Only required for a New Version summary, otherwise leave out.
{\em Summary of revisions:}*\\
  %Only required for a New Version summary, otherwise leave out.
{\em Nature of problem: high--energy collisions of heavy nuclei with protons or
each other, are used to study the influence of a Quark--Gluon Plasma on particle 
production mechanisms. The relation between calculations of such effects, and the 
complex underlying final state is poorly understood, making correct comparison
between theory and data difficult.}\\
  %Describe the nature of the problem here. \\
{\em Solution method: the underlying event is generated in a multi-step procedure,
relying on a calculation of collision geometry, as well as input from a proton--proton
Monte Carlo event generator, run by the user. This allows the user to pick and choose 
between different well-understood proton--proton models for the underlying event. 
Secondly, the signal process can be supplied either from a proton--proton generator,
or from a generator implementing interactions with a Quark--Gluon Plasma.}\\
  %Describe the method solution here.
{\em Additional comments including Restrictions and Unusual features: The user must 
install the input event generator(s), which must be able to output events in the 
standard HepMC event format [1].}\\
  %Provide any additional comments here.
   \\

\end{small}

\section{Introduction} \label{sec:intro} Hadron collisions are complex processes
involving many different phenomena from the soft and the electroweak sector,
some calculable by perturbative techniques, and some -- especially in the case
of soft QCD processes -- relies on the use of models. Over the past 20 years,
such techniques have been refined and implemented into so-called General Purpose
Monte Carlo Event Generators, such as  Herwig~7~\cite{Bellm:2015jjp}, 
Pythia~8~\cite{Sjostrand:2014zea} and Sherpa~\cite{Gleisberg:2008ta}, which provide a very
good description of most aspects of hadronic collisions. The connection between
these well-understood models of hadronic collisions, and collisions of heavy
nuclei are, however, lacking.

Existing event generators for the study of jets in heavy ion collisions are
roughly speaking divided into two categories. In the first category, there are
generators which modify a proton--proton (pp) simulation according to a set of assumptions
about the influence of a Quark-Gluon Plasma on the jet evolution. Generators
such as JEWEL~\cite{Zapp:2013vla} and the JETSCAPE generators~\cite{Cao:2017zih}
place themselves in this category. On the other hand, generators such as HIJING
\cite{Wang:1991hta} or the recent Angantyr addition to Pythia~8
\cite{Bierlich:2016smv,Bierlich:2018xfw} aim to describe the full event. Generators of the latter
type are generally better suited to investigate the influence of the heavy ion
"underlying event", \textit{i.e.} the presence of several jets from almost
uncorrelated sub-collisions, disturbing the jet signal of interest. In spite of these
generators' ability to generate full events, a user is still faced with a
problem given the task of (a) generating a jet using a given standard tool (with
or without the assumption of a plasma) (b) embedding this into a realistic
underlying event. Though a seemingly trivial task, it is nevertheless important
to carry this task out with sufficient rigour; in particular, one cannot assume
that an underlying event consists of a number of uncorrelated pp
collisions, as one will at least have to respect energy--momentum conservation,
restricting the phase space of the underlying event. It is also important for
a more technical reason. When comparing heavy ion calculations to data, the
comparison is usually carried out at various "centralities", understood
as the impact parameter at which the nuclei have collided. Since the impact
parameter is not experimentally accessible, experiments use a different definition,
usually based on event activity. Therefore a direct comparison is at least difficult.
With a full simulation of the underlying event, the direct comparison is 
enabled by construction.

The present program introduces a modified implementation of the underlying event model
first introduced in the Fritiof event generator~\cite{NilssonAlmqvist:1986rx,Pi:1992ug},
and recently implemented in Pythia~8~\cite{Bierlich:2016smv,Bierlich:2018xfw}, 
with the caveat
that this implementation allows for a given HepMC~\cite{Dobbs:2001ck} compatible {\it{pp}} event 
generator to produce the underlying event, and a given HepMC compatible generator to
produce the signal event. This allows the user to perform a quantitative
simulation study of jets in a semi-realistic heavy ion background. 

\section{Structure of the program} \label{sec:struct} 

For the program to work,
input from one or more event generator(s) supporting the HepMC
 event standard is needed. The event generator input is read
from FIFO--pipes and an initial state calculation determines how many events
should be taken from each pipe, and used to build up a full proton--Nucleus ({\it{pA}})  or
Nucleus--Nucleus ({\it{AA}})  event.  The program is written in Python, and structured in two
programmatic parts.  

\begin{enumerate} 
\item An initial state calculation
determining how many sub-collisions a given event should consist of.  
\item A simulation step taking input from the event generators, and piping it, according
to the initial state calculation, into a full {\it{pA}} or {\it{AA}} event.  
\end{enumerate}

\begin{figure} 
\begin{center} 
\includegraphics[width=\textwidth]{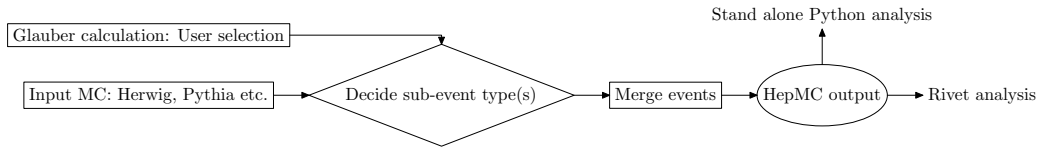}
\end{center} 
\caption{\label{fig:flowchart}Flowchart indicating the typical workflow. 
The user initializes a Glauber calculation along with the input
generator(s), the events are merged together, and a single heavy ion event is
produced in the HepMC format.} 
\end{figure}

The program workflow contains several logic parts, as indicated in the flowchart 
in figure \ref{fig:flowchart}. A Glauber calculation is performed, and
serves, along with the event generator output, the input to the decision process,
determining how many events and which types of events should be merged into a
heavy ion event. The output is a HepMC event with an associated weight, which
the user can perform an independent analysis of inside the Python framework
itself or pipe to an external analysis framework such as Rivet~\cite{Buckley:2010ar}.
We highly recommend the user to take advantage of the latter option.

Before running the program, the user should download and compile the event
generator(s) they desire to use. The program comes with sample run cards for
Herwig~7, as well as sample programs for Pythia~8, ready to use with the default
installation of the two event generators.  The program
itself is written in Python, and is highly customizable. In the main script,
the user should set paths to event generator supplied FIFO--pipes, as well as
basic run information related to the colliding nuclei.

\section{Glauber calculation} \label{sec:glauber} 

To determine the number of sub-collisions needed to generate the full heavy ion 
collision, a so--called Glauber~\cite{Glauber:1955qq} calculation is performed. 
Nucleons are placed in a Woods-Saxon potential in a stochastic manner, nuclei 
from projectile and target which are overlapping in impact-parameter space, are 
counted as "collided" in a given event.

The user must supply the Glauber calculation with the nuclear PDG number (on the
form 100ZZZAAAI for nuclei, 2212 for protons) as well as the inelastic,
non--diffractive cross section for a nucleon--nucleon collision ($\sigma_{NN}$)
at the desired collision energy.

The Glauber calculation can in principle be run stand-alone, should a user
desire to do that. In such cases, the calculation should be initialized in a
similar way and information about the position of individual sub--collisions can be
accessed by calling the \texttt{next} method of the initialized Glauber object.
This returns a list of particles and sub--collision, which can be examined
further.

\subsection{Sampling the Woods-Saxon distribution} The nucleus' transverse
structure is described by a Woods-Saxon distribution. We use here the
parametrization by Broniowski \textit{et al.} known as the GLISSANDO
parametrization~\cite{Broniowski:2007nz,Rybczynski:2013yba} with a density function:

\begin{equation} 
\rho(r) = \frac{\rho_0(1+wr^2/R^2)}{1 + \exp((r - R)/a)}.
\end{equation} 

Here $R$ is the nuclear radius, $a$ is the nuclear skin width,
$\rho_0$ is the central density (not used in the present calculation) and $w$
the Fermi parameter. 

\subsection{Performing sub-collisions} 

\begin{figure} 
\includegraphics[width=0.5\textwidth]{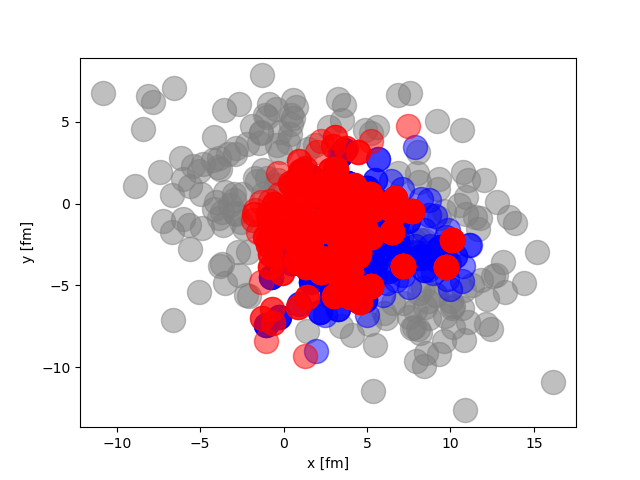}
\includegraphics[width=0.5\textwidth]{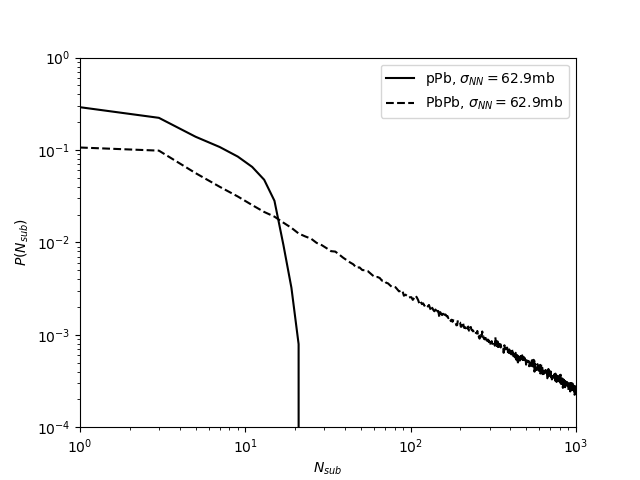}
\caption{\label{fig:wounded}(left) Sample {\it{PbPb}}-event at intermediate impact
parameter, showing overlapping nucleons leading to sub-collisions. (right) The
distribution of number of sub-collisions in {\it{PbPb}} and {\it{pPb}} with $\sigma_{NN} =
62.9$mb, indicating the vast difference in complexity for the two types of
systems.} 
\end{figure}

The nuclear impact parameter is chosen
from a Gaussian distribution, and in the default mode, the two nuclei are
collided using the simplest possible recipe, namely by treating each nucleon as
a black disk with radius $\sqrt{\sigma_{NN}/4\pi}$. If two nuclei are
overlapping, they collide. In figure \ref{fig:wounded} (left) a sample {\it{PbPb}}
collision is shown. Grey nucleons do not participate, while the participant
nucleons from target and projectile are coloured red and blue respectively. The
geometry of an event is chosen such that the target sits at coordinates (0,0),
while the projectile is shifted by the impact parameter as well as an event
plane angle.

The physical value for $\sigma_{NN}$ is the only free parameter. The user can 
either insert a value as desired or rely on a calculated value. The program 
includes a simple implementation of the Schuler--Sj{\"o}strand 
model~\cite{Schuler:1993wr} which, given the center--of--mass energy, outputs
the total, elastic and single diffractive cross sections, calculated
using a fairly standard Pomeron--based scheme. The user may prefer to use 
experimentally obtained values instead, and rely on the Schuler--Sj{\"o}strand
model only in cases where measurements are not available.

For the purpose of studying the background to jets in heavy ion collisions, such
a simple treatment ought to be enough. The program is, however, devised in such
a way that the user can easily extend to other, more sophisticated models for
sub-collisions. The basic functionality is implemented in a class called
\texttt{GlauberBase}, which for the case of black--disk--collisions, is derived
from a class called \texttt{BlackDisk}, which is the class instantiated by the
user. Should a user wish to collide for example semi-transparent disks, such a
class can be added by the user, inheriting from the same base class. Thus all
basic functionality is retained, and the user needs only implement the
functionality connected to the semi-transparent disk. 

In figure \ref{fig:wounded} (right), the distribution of number of sub--collisions
for a sample {\it{pPb}} system and a sample {\it{PbPb}} system is shown. Note that the
$x$-axis is logarithmic, indicating that the computational complexity for
treating a {\it{PbPb}} system can be orders of magnitudes larger than a {\it{pPb}} system. The
reason being that each projectile nucleon has the possibility to interact with
several target nucleons, none being limited to only one interaction.

\subsection{Glauber--Gribov fluctuations}\label{sec:gg-fluct}

In asymmetric {\it{pA}} collisions, it is of key importance to include colour fluctuations
of the projectile, allowing $\sigma_{NN}$ to fluctuate event by event, to get a
realistic estimate of the number of sub-collisions~\cite{Alvioli:2013vk}. 
For this purpose a \texttt{GlauberGribov} model has been implemented as an extension
to \texttt{GlauberBase}, where $\sigma_{NN}$ is drawn from a log--normal distribution~\cite{Bierlich:2016smv}:
\begin{equation}
	P(\ln (\sigma_{tot})) = \frac{1}{\Omega \sqrt{2\pi}} \exp \left(\frac{\ln^2(\sigma_{tot}/\sigma_0)}{2\Omega^2} \right), \sigma_{NN} = \lambda \sigma_{tot}.
\end{equation}
Parameters suitable for 5.02 TeV collisions, $\Omega = 0.25, \sigma_0 = 85, \lambda = 0.63$
are set by default.

\section{Merging sub-events} \label{sec:merge}
The backbone of the Angantyr merging recipe is the notion of "secondary absorptive"
sub--collisions. The concept is most easily explained in the simple case of a proton--Deuteron ({\it{pD}})
collision. Consider the case where the proton has colour exchange 
with both Deuteron constituents. Clearly,at least one colour exchange must exist over the 
full kinematically allowed rapidity span of the event. 
Further exchanges do not have to but can span only a part of this. In {\it{pp}} event generators 
various so--called "colour reconnection" models lead to such effects. In the {\it{pD}} collision, the 
same mechanism is mimicked by labelling one of the two sub--collisions as the "primary" one, 
and the other as the "secondary" one. The primary one will contribute to the final state as a 
normal {\it{pp}} collision (including colour reconnection), but the secondary one does not need to 
have even a single colour exchange over the full kinematically allowed region.
It is observed~\cite{Bierlich:2018xfw} that such secondary absorptive events contributes approximately 
like a single diffractive event, and will thus be modelled as such. 

For a full explanation of this procedure, and its generalization to {\it{AA}} collisions, we 
refer to ref.~\cite{Bierlich:2018xfw}.

\subsection{Combination of events} 

Even given the recipe presented for combining sub-events presented in Angantyr,
one is left with several choices regarding specific implementation. 
In the current program the sub-events are combined as follows: 
\begin{enumerate} 
\item Sort possible collisions from Glauber calculations according to their impact
parameter (most central first).  
\item If none of the participants was used in a "harder" (defined below) event, 
this sub-collision is labelled "hard".  
\item If {\it{one}} of the participants was 'used' before, its labelled
as additional\_{\it{side}}, with {\it{side}} being the direction the participant
was used in a previous harder collision.  The other participant is then also
labelled as used and therefore removed from possible hard processes.  
\item If both participants have been used before in harder collisions, we label the
sub-collision as additional\_both.  
\end{enumerate} 

Here, the measure of {\it{hardness}} is calculated in the {\tt selectHardest} function. 
The user can select between several hardness measures, or define their own. The default
supplied (and recommended) defines hardness as the scalar sum of hadron $p_\perp$ 
(dubbed \texttt{HT} in the program). The implementation details of how to fill the 
list according to hardness are outlined in  \ref{app:correlatedCollisions}.
For example in a {\it{pA}} collision, this will produce one 'hard' event and since the proton 
was then 'used', the additional
sub-collisions are added to the nucleus side. Once the labelling is performed,
the main event is defined by the hardest sub-collision.  In the spirit of ref.~
\cite{Bierlich:2016smv} the main event and the following hard sub-collisions
are chosen to be the {\it{hardest}} of a $N_{sub}$ of  absorptive QCD events.

\subsection{Momentum conservation}

In order to include momentum conservation we employ a very primitive scheme 
to restrict the momentum in beam directions to not exceed the incoming energies.
Since events produced by an arbitrary generator are used, the program should not rely on event record information.
Instead we keep track of the sum of incoming momentum in $z$-direction for each participant and add for each
stacked event the sum of final state particles in a given rapidity range $[0,\eta]$ and 
$[-\eta, 0]$ to the corresponding incoming participant. If the next event exceeds the allowed incoming
momentum we do not add the corresponding event.

The default cut-off value is set to $\eta = 7.0$, which includes the acceptance of 
all present detector experiments, but excludes beam remnants. The user can change this value if desired.

\subsection{Adding a signal process} 

It is possible to add a signal process in the sub--collision with smallest impact parameter,
by generating the process using the input generator to
create another FIFO--pipe containing those events. As indicated in the
introduction, we foresee the most common use case of the program will be to add
a signal event such as $Z+$jet(s) or di--jets with or without the assumption of a medium,
and use the program to study the behaviour of
the jet in a heavy ion background. If one wishes to have the correct cross section
for such an event, the event should be reweighted with a factor
$N_{sub}\sigma_{signal}/\sigma_{NN}$, where $N_{sub}$ is the number of
sub-collisions.

Since different general purpose event generators have different models for QCD
MPI events and different Matrix Element + Parton shower matching and merging
schemes implemented, it might be beneficial to take the heavy ion background
from one generator, and adding a signal from another generator. This can be done
directly by setting up FIFO--pipes from the separate generators.

It is also possible to add a signal process from an event generator which includes a medium 
response, such as JEWEL or a JETSCAPE event generator. This will enable a study of the medium
modified jet in a realistic background simulation.

\subsection{Tuning}

General purpose MC generators include several parameters for describing multi parton interactions, 
defining the underlying event to a signal process in pp. If these parameters are changed, the model, 
of course, loses its predictive power for minimum bias heavy ion observables, but if the purpose is
to provide modelling of the background to a signal process, this predictive power can be traded off 
for a better description of the underlying event. In such cases, the user might want to retune these
parameters. Users are referred to user manuals for input generators to carry out this process, we 
provide two sample Rivet analyses for {\it{pA}}, and one for {\it{AA}}, to do such a retuning.
The sample results in the next section are all using default tunes.

\section{Reading out results} \label{sec:results}

Results are written out in HepMC format, enabling easy comparisons to data using the Rivet
framework, or can be analysed directly by the user. In this section we 
demonstrate the program functionality by showing comparisons of minimum bias {\it{pPb}} and {\it{PbPb}} 
data to Pista + Pythia~8 and Pista + Herwig~7, but remind that any {\it{pp}} generator able 
to fulfil the requirements listed in section \ref{sec:merge}, and can produce HepMC output 
files can be substituted for those.
Secondly, we demonstrate sample results obtained by embedding a $Z$+jet events into {\it{PbPb}} 
collisions. We demonstrate using first Pythia~8 and Herwig~7 signal events, following up with events
produced using JEWEL~\cite{KunnawalkamElayavalli:2016ttl}. But also here any generator able to produce HepMC events can be used.

The Rivet comparison routines used here are shipped with the program to allow the user
to reproduce all plots shown in this paper directly.

\subsection{Minimum bias results}\label{sec:minbias}

We present here comparisons to multiplicity distributions obtained with the ATLAS and ALICE experiments
at the LHC in the case of {\it{pPb}}~\cite{Aad:2015zza} and {\it{PbPb}}~\cite{Aamodt:2010cz,Aad:2015wga} collisions. 
This serves the purpose of demonstrating that for the case of minimum bias collisions, the model does well in
estimating the total particle production.
Thus the model provides a sensible baseline for modelling the underlying event to a jet.

\begin{figure}
	\begin{center}
		\includegraphics[width=0.65\textwidth]{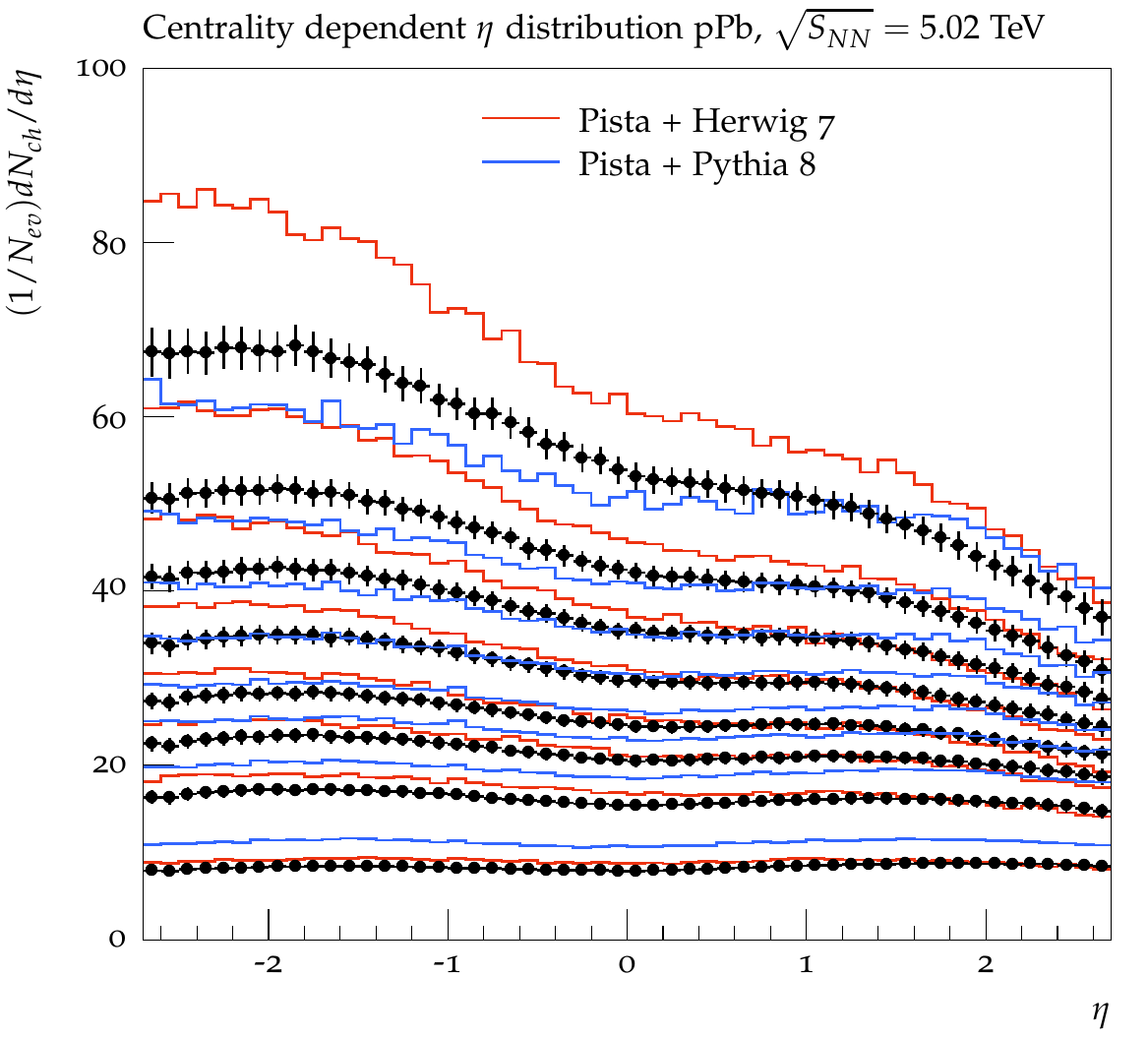}
	\end{center}
	\caption{\label{fig:pa-mult}Centrality dependent $\eta$ distributions of charged primary
	particles in {\it{pPb}} collisions at $\sqrt{s_{NN}} = 5.02$ TeV, obtained by ATLAS~\cite{Aad:2015zza}.
	Centralities ranging from top to botton are: 0-1\%, 1-5\%,...,60-90\%. Experimental results are
	compared to Pista + Pythia~8 and Pista + Herwig~7.}
\end{figure}

The centrality measure used for comparison is the same as (or similar to) the ones used in the experimental
analyses. In the case of ATLAS analyses, the centrality measure is defined as the $\sum E_\perp$ of particles
in $2.09 < |\eta| < 3.84$ (only the {\it{Pb}} going side for {\it{pPb}}), which can be directly applied. For ALICE analyses
no direct, particle level observable exist, as the centrality observable is not detector unfolded. We use, as
a proxy for the centrality observable, $\sum E_\perp$ of charged particles in the relevant detector acceptance.
This, as well as the implementation of experimental cuts, can be changed as desired.

As mentioned in section \ref{sec:gg-fluct}, Glauber-Gribov fluctuations are necessary to reproduce the 
centrality observable in {\it{pPb}} collisions, and is thus employed for those results. We will not go further 
into the physics details of this point here, but it would be well worth to further study the effects of such
fluctuations on final state observables.

In figure \ref{fig:pa-mult} we show multiplicity in {\it{pPb}} collisions as a function of pseudo-rapidity,
in bins of centrality as defined above. 
While the Pista + Herwig~7 description (red line) visibly overshoots most of the distributions, 
the Pista + Pythia~8 (blue line) simulation undershoots the very central event multiplicity 
and overshoots more peripheral event multiplicities. 

While the tilt in the multiplicity distributions  in the Herwig~7 description 
is slightly too steep w.r.t. data, the Pythia~8 comparison shows an opposite 
less tilted behaviour w.r.t. data. 

It is worth mentioning that a better agreement can be achieved by independent retuning 
the parameters used in the Pista implementation for the two generators. 
In this contribution, we limit ourselves to show distributions with equal parameters 
on the Pista part and default parameter settings on the generator side.  

\begin{figure}
	\begin{center}
		\includegraphics[width=0.65\textwidth]{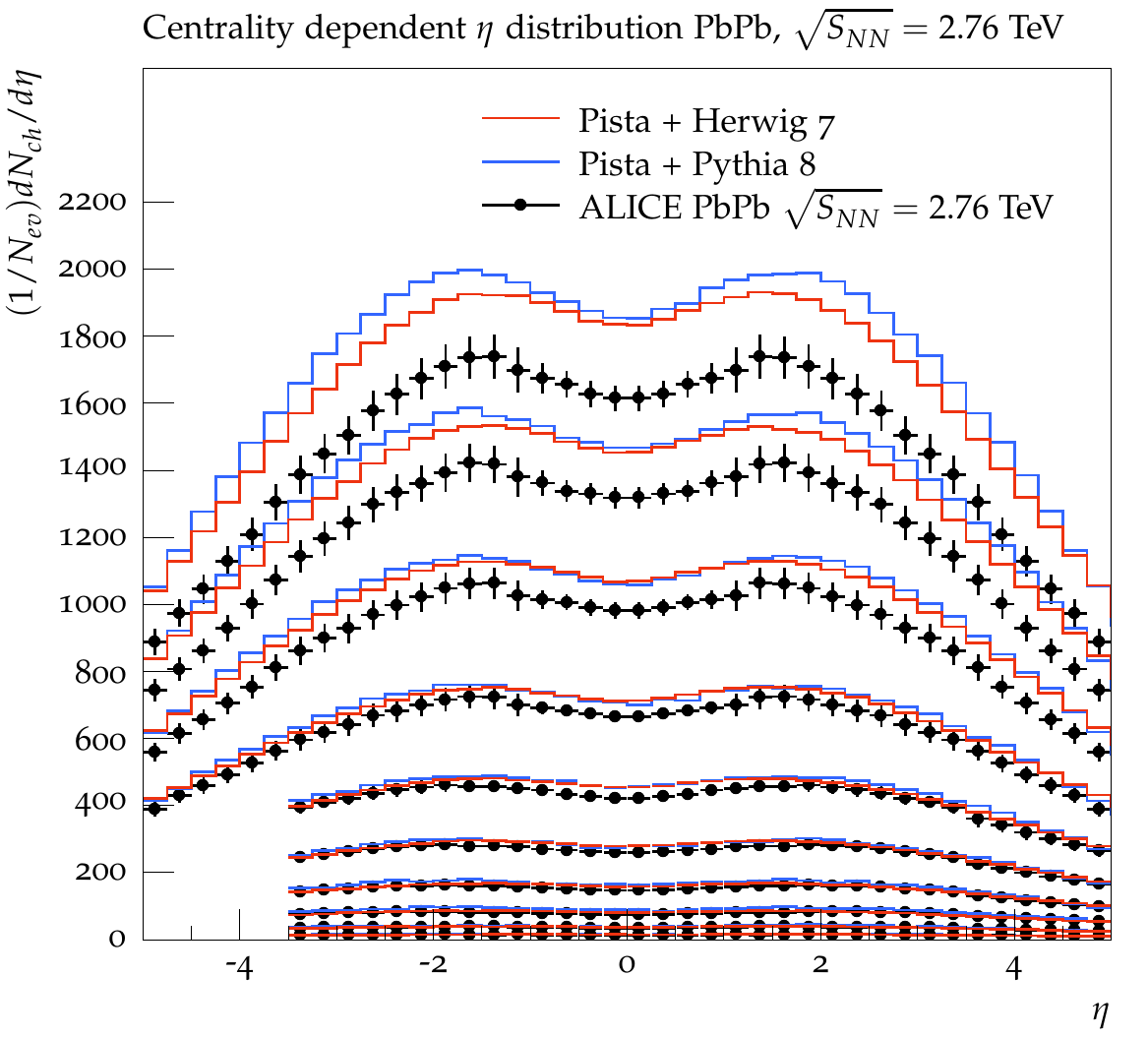}
	\end{center}
	\caption{\label{fig:aa-mult}Centrality dependent $\eta$ distributions of charged primary particles 
	in {\it{PbPb}} collisions at $\sqrt{s_{NN}} = 2.76$ TeV, obtained by ALICE~\cite{Aamodt:2010cz}. Centralities
	ranging from top to botton are: 0-5\%, 5-10\%,...,80-90\%. Experimental results are compared to 
	Pista + Pythia~8 and Pista + Herwig~7.}
\end{figure}

In figure \ref{fig:aa-mult} we show the multiplicity of charged particles in Pb-Pb collisions as
a function of pseudo-rapidity in bins of centrality.
Both the generators overshoot the data for very central events by approximately 
20\% and show a better agreement for lower more peripheral collisions. 
Compared to the {\it{pPb}} comparison we observe a less pronounced discrepancy between 
the two generators. 
Compared to other models for heavy ion collisions, the description in this simple model 
is slightly worse than in the evolved Angantyr model implemented in Pythia~8, but similar
to the HIJING~\cite{Wang:1991hta} and AMPT~\cite{Lin:2004en} models. (For comparison to
the latter models, see ref.~\cite{Aamodt:2010cz}.) We do not carry out any detailed comparison
to other models in this paper, as this is beyond the scope of a manual, but defer it to 
a coming publication.

\begin{figure}
	\begin{center}
		\includegraphics[width=0.45\textwidth]{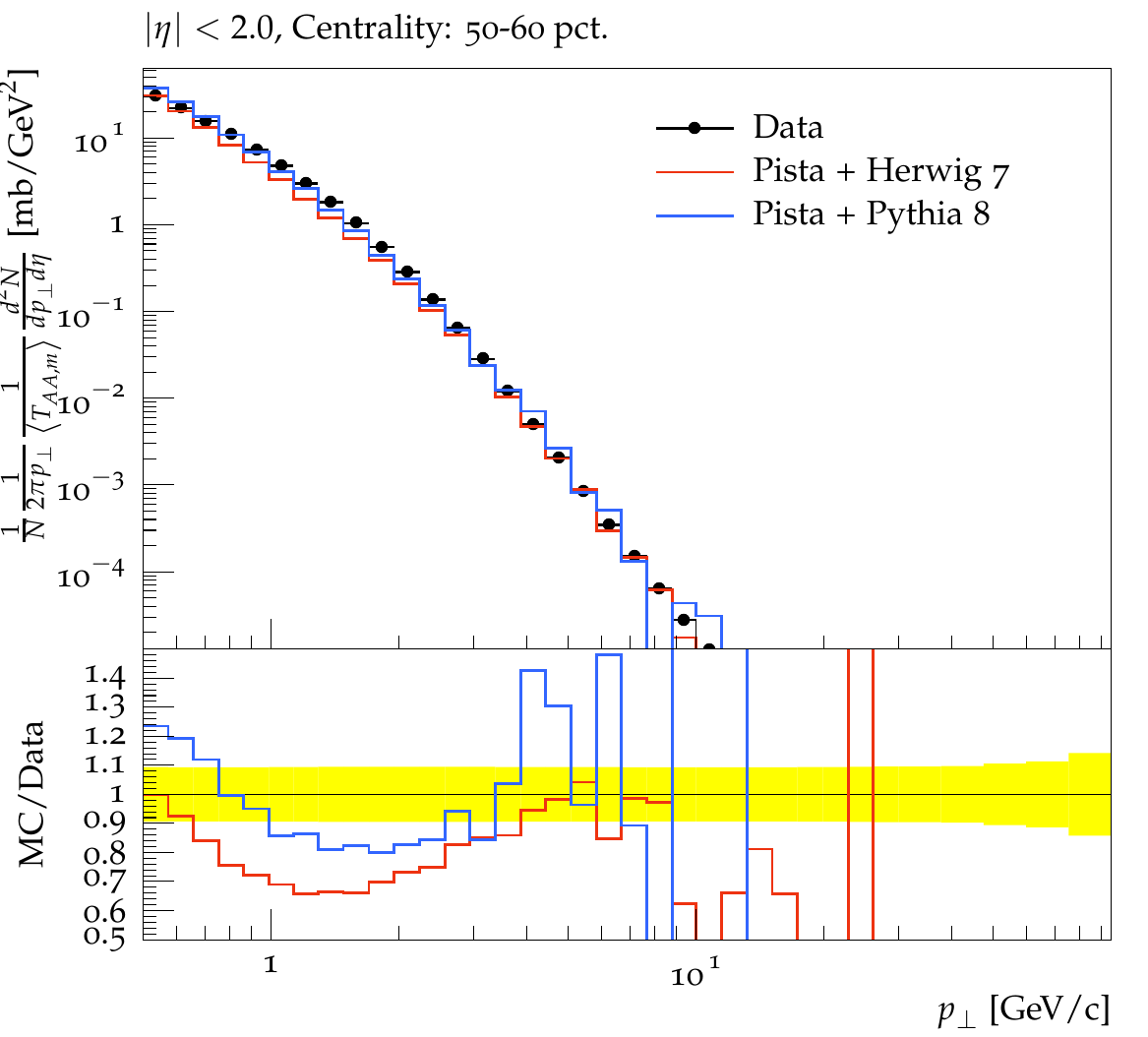}
		\includegraphics[width=0.45\textwidth]{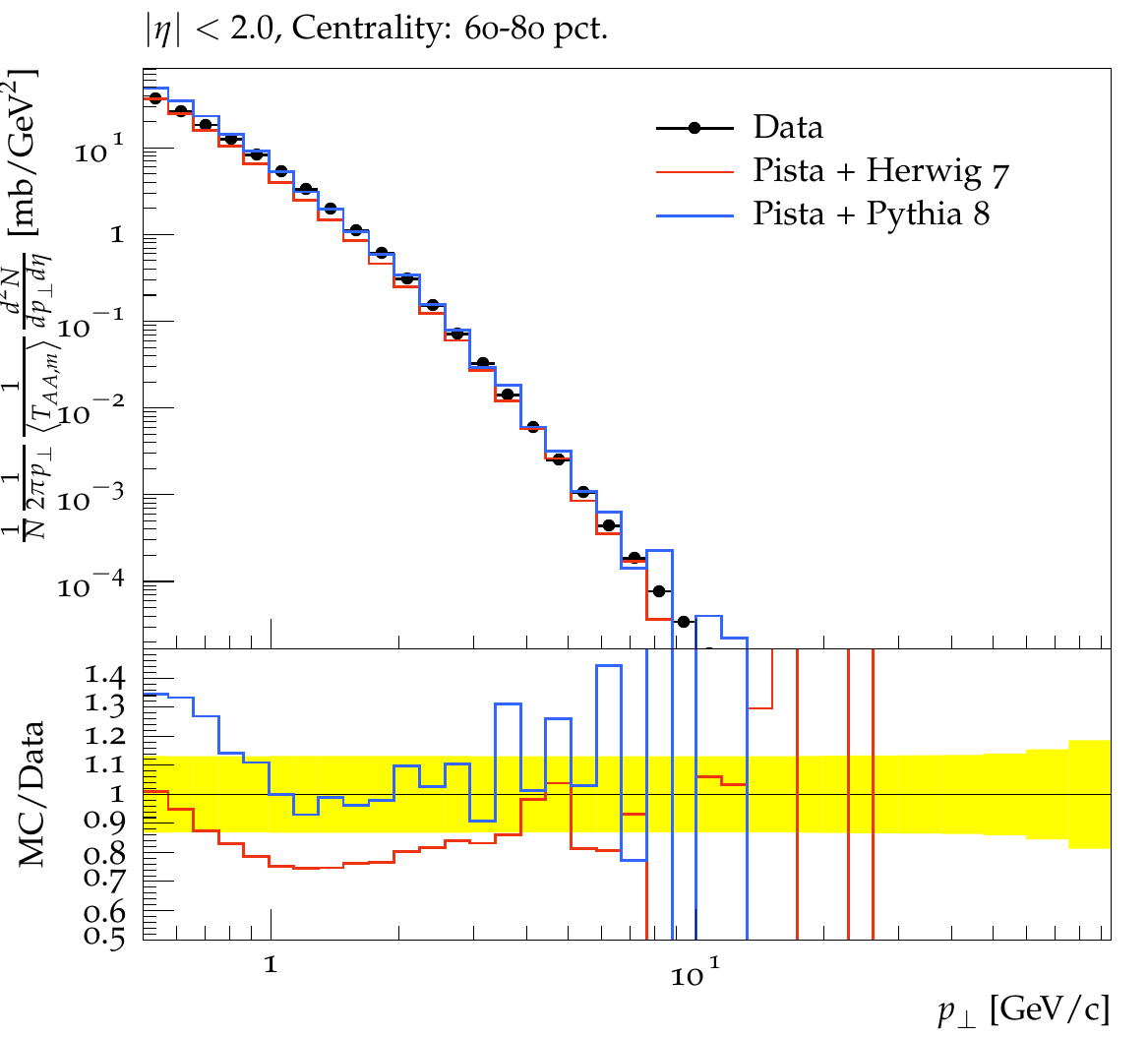}
	\end{center}
	\caption{\label{fig:aa-pt1}Centrality dependent $p_\perp$ distributions in {\it{PbPb}} collision at
	$\sqrt{s_{NN}} = 2.76$ TeV, obtained by the ATLAS experiment~\cite{Aad:2015wga}. Mid--central
	(50-60\%) left and peripheral (60-80\%) right.}
\end{figure}

\begin{figure}
	\begin{center}
		\includegraphics[width=0.45\textwidth]{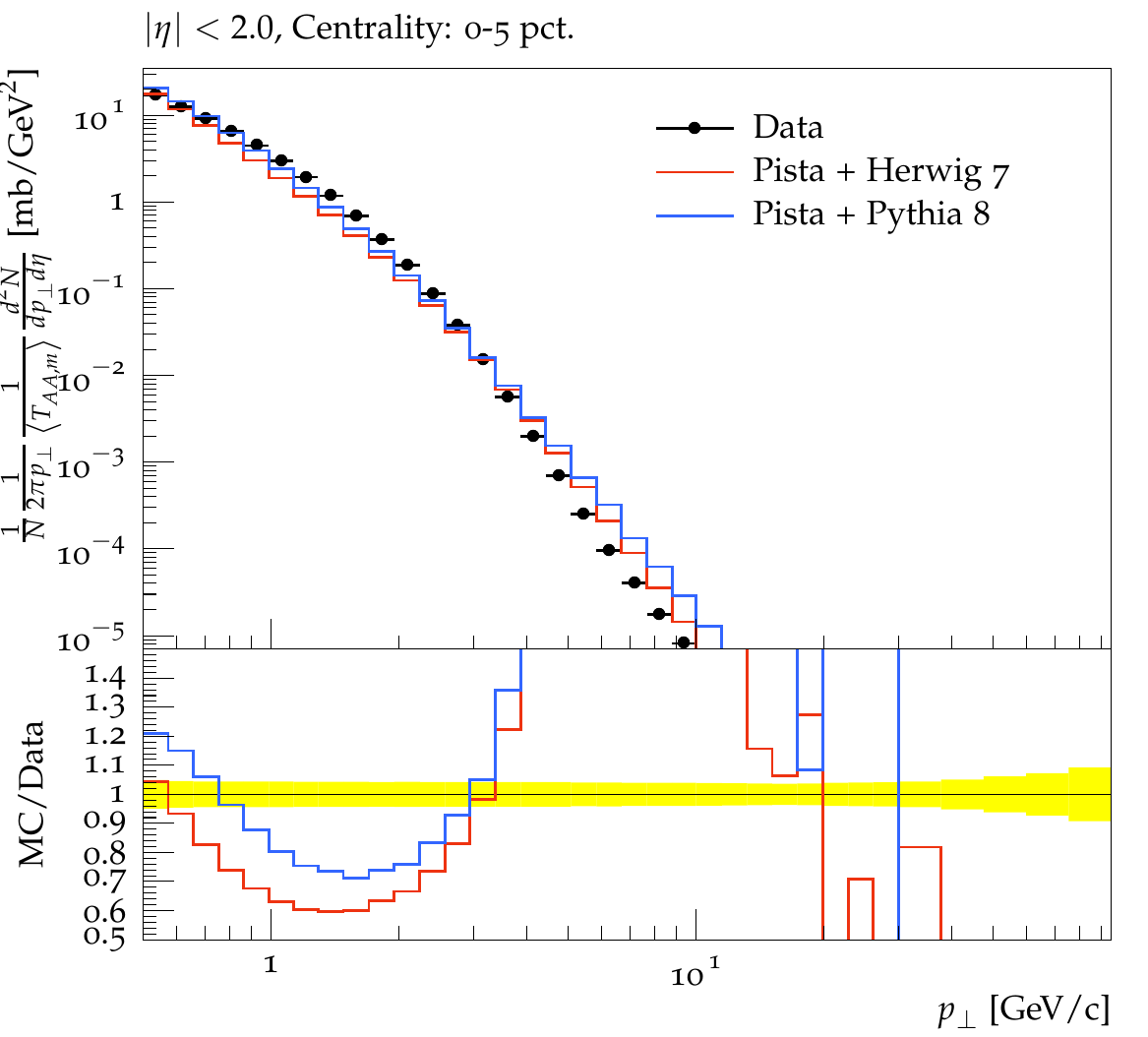}
		\includegraphics[width=0.45\textwidth]{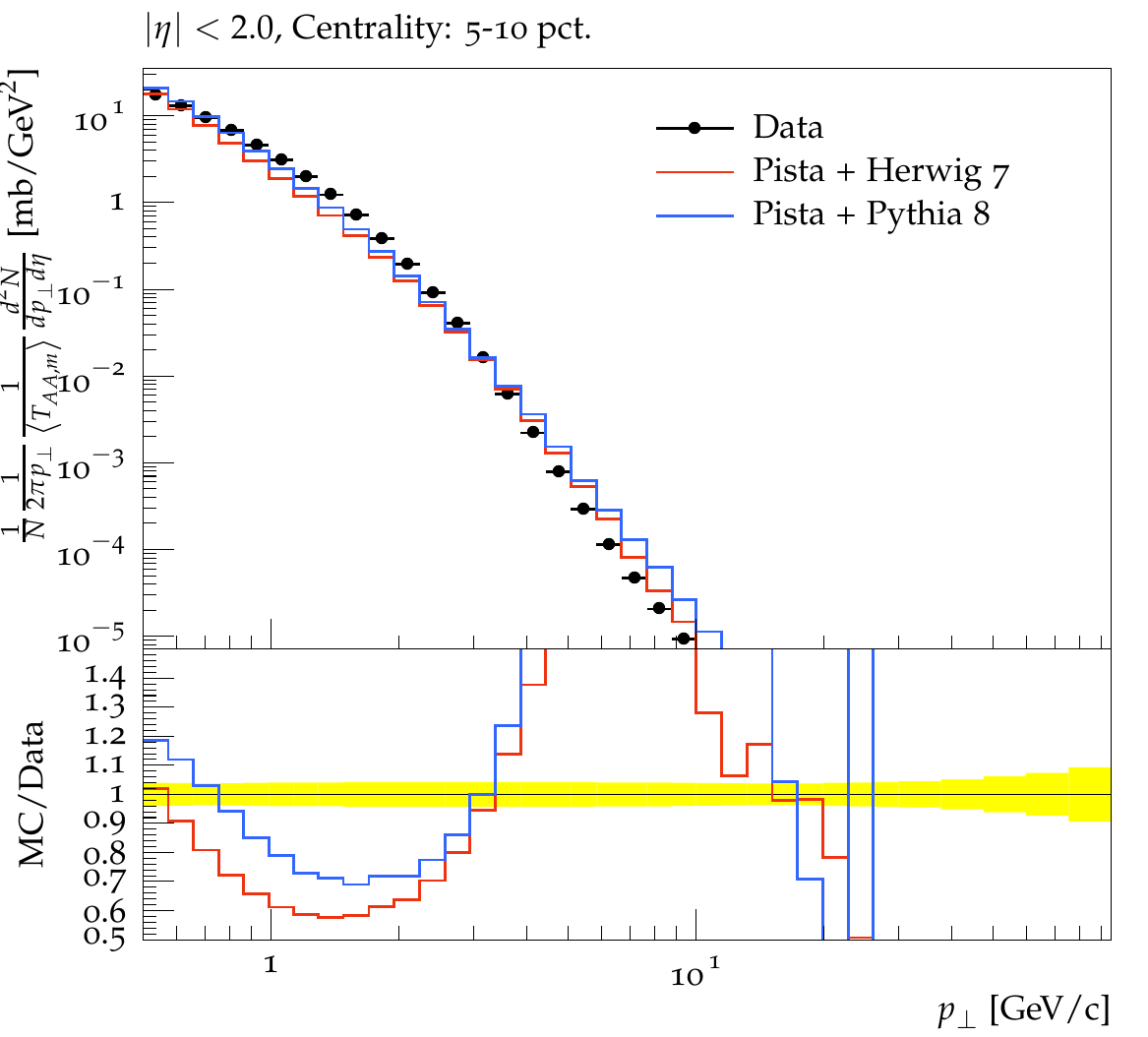}
	\end{center}
	\caption{\label{fig:aa-pt2}Centrality dependent $p_\perp$ distributions in {\it{PbPb}} collision at 
	$\sqrt{s_{NN}} = 2.76$ TeV, obtained by the ATLAS experiment~\cite{Aad:2015wga}. Most central (0-5\%) left and 5-10\% right.}
\end{figure}

In figures \ref{fig:aa-pt1} and \ref{fig:aa-pt2} we compare to $p_\perp$ distributions in {\it{PbPb}} collisions
obtained by ATLAS~\cite{Aad:2015wga}. 
The peripheral and semi-peripheral distributions in figure \ref{fig:aa-pt1} are fairly described by both 
generators, Pista + Herwig~7 is systematically below Pista + Pythia~8, an investigation whether this is 
due to differences in the parton shower implementation or hadronization models in the generators would be
interesting from a physics perspective.

The corresponding central distributions in figure \ref{fig:aa-pt2} are described worse, similarly to
the other previously mentioned event generators which does not include medium effects.

\subsection{Results with an embedded jet}

We now go on to demonstrate the addition of signal processes in the case of $Z$+jet events embedded in a
{\it{PbPb}} background.

We provide not a full physics analysis but defer the discussion of such to a coming paper. We instead 
demonstrate the features by showing simple features of events with a $Z\rightarrow \mu^+ \mu^-$ plus a
jet reconstructed by the anti-kT algorithm implemented in FastJet~\cite{Cacciari:2011ma}. A jet is said
to be back-to-back with the $Z$ (reconstructed from the muon pair) if it has $\Delta \phi > 7\pi/8$ with
the reconstructed $Z$. The $Z$ is required to have $p_\perp > 60$ GeV, and the muons $p_\perp > $10 GeV. 

\begin{figure}
	\begin{center}
		\includegraphics[width=0.45\textwidth]{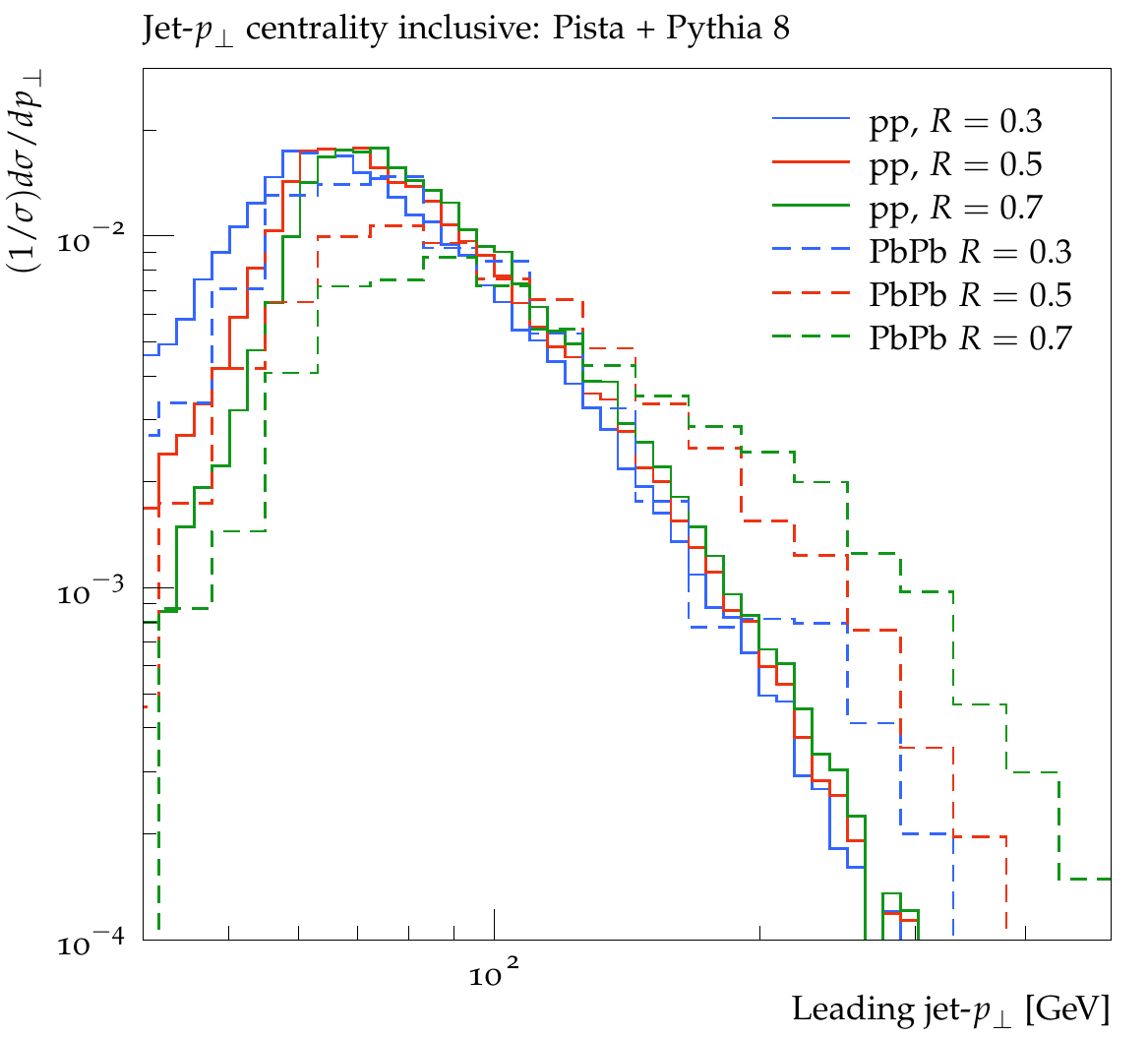}
		\includegraphics[width=0.45\textwidth]{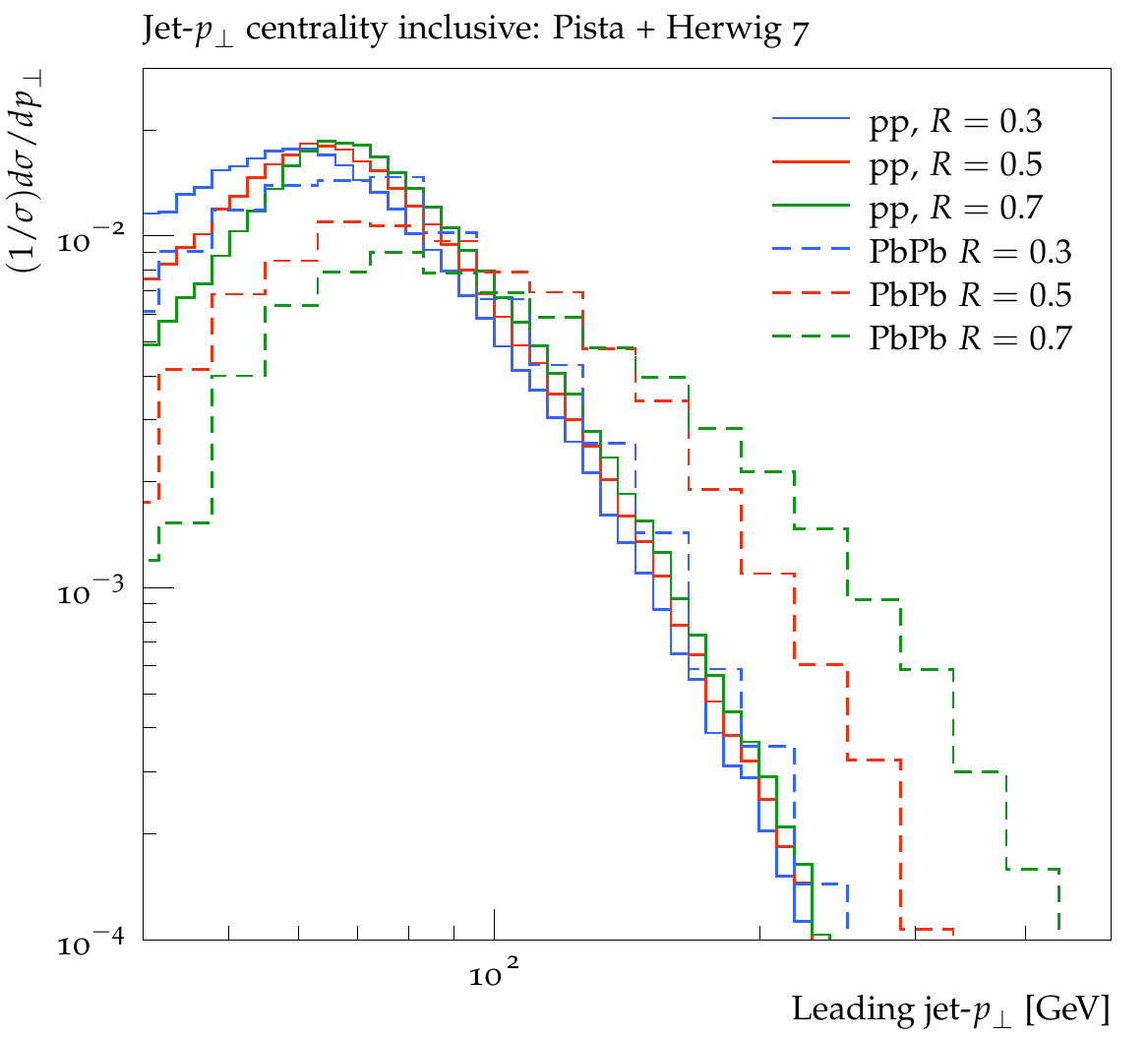}
	\end{center}
	\caption{\label{fig:jetpt-inc}Transverse momentum of the leading jet for a $Z+$ jet event embedded
	into a {\it{PbPb}} event, using Pythia~8 (left) and Herwig~7 (right) respectively. In the figures jet cone
	radii of $\Delta R = 0.3, 0.5$ and $0.7$ are shown, and also compared to {\it{pp}} (\textit{i.e.} the embedded signal only).}
\end{figure}

In figure \ref{fig:jetpt-inc} we show the jet-$p_\perp$ of the leading jet in the event for an embedded
event, compared to the embedded {\it{pp}} event only, for a jet cone radius of $\Delta R = 0.3, 0.5$ and $0.7$.
For comparison, we show Pista + Pythia~8 (left) and Pista + Herwig~7 (right). It is is directly visible
that the chosen $\Delta R$ parameter has significantly higher influence in a {\it{PbPb}} embedded event than in
a {\it{pp}} event, which should not be surprising as a {\it{PbPb}} event has much more underlying event activity, swept
up by the jet algorithm. We note also that choosing a sufficiently narrow jet cone in {\it{PbPb}} makes the jet
coincide with the {\it{pp}} jet. The Pista + Pythia~8 and Pista + Herwig~7 figures are very similar, as one would expect.

\begin{figure}
	\begin{center}
		\includegraphics[width=0.45\textwidth]{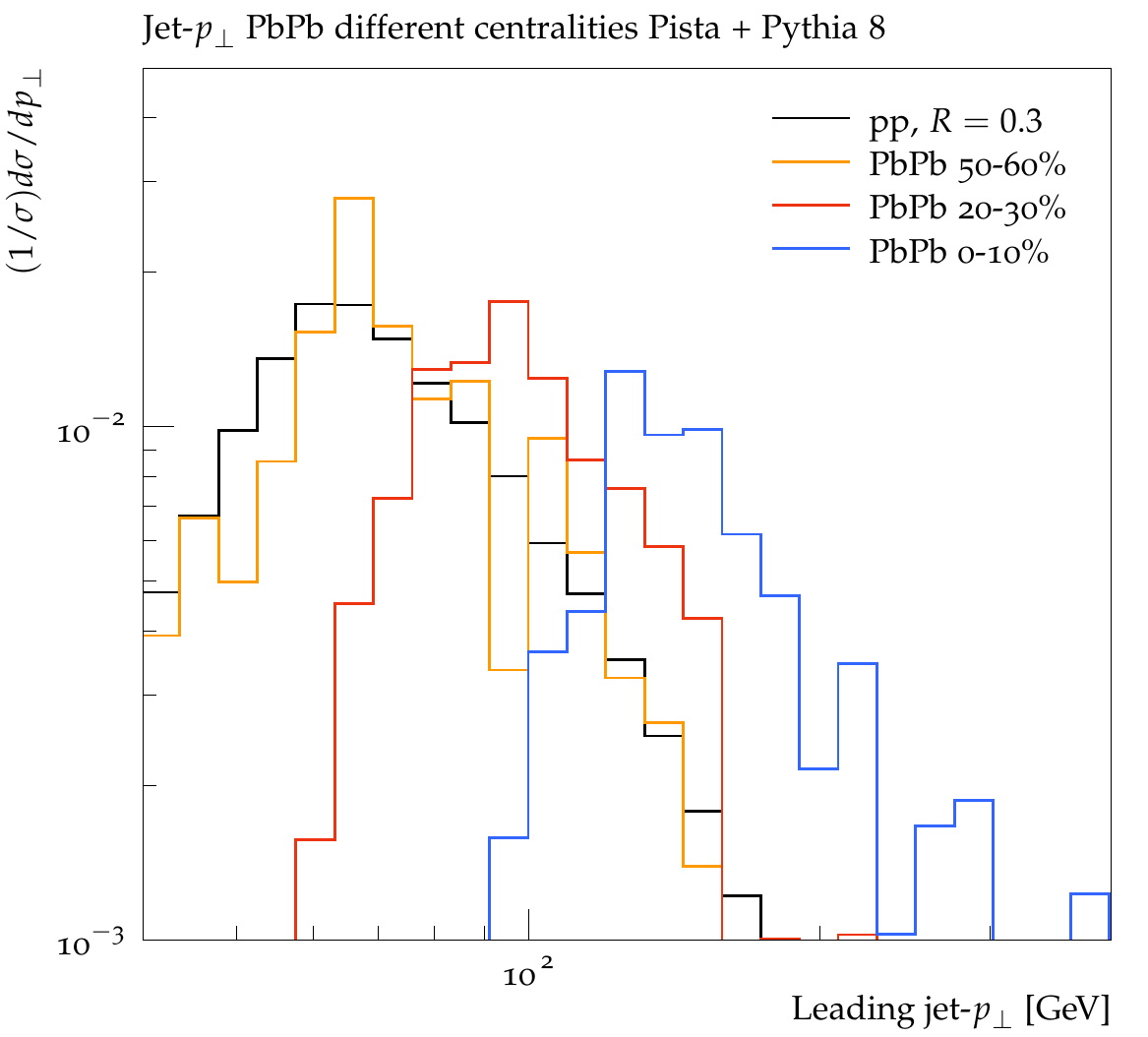}
		\includegraphics[width=0.45\textwidth]{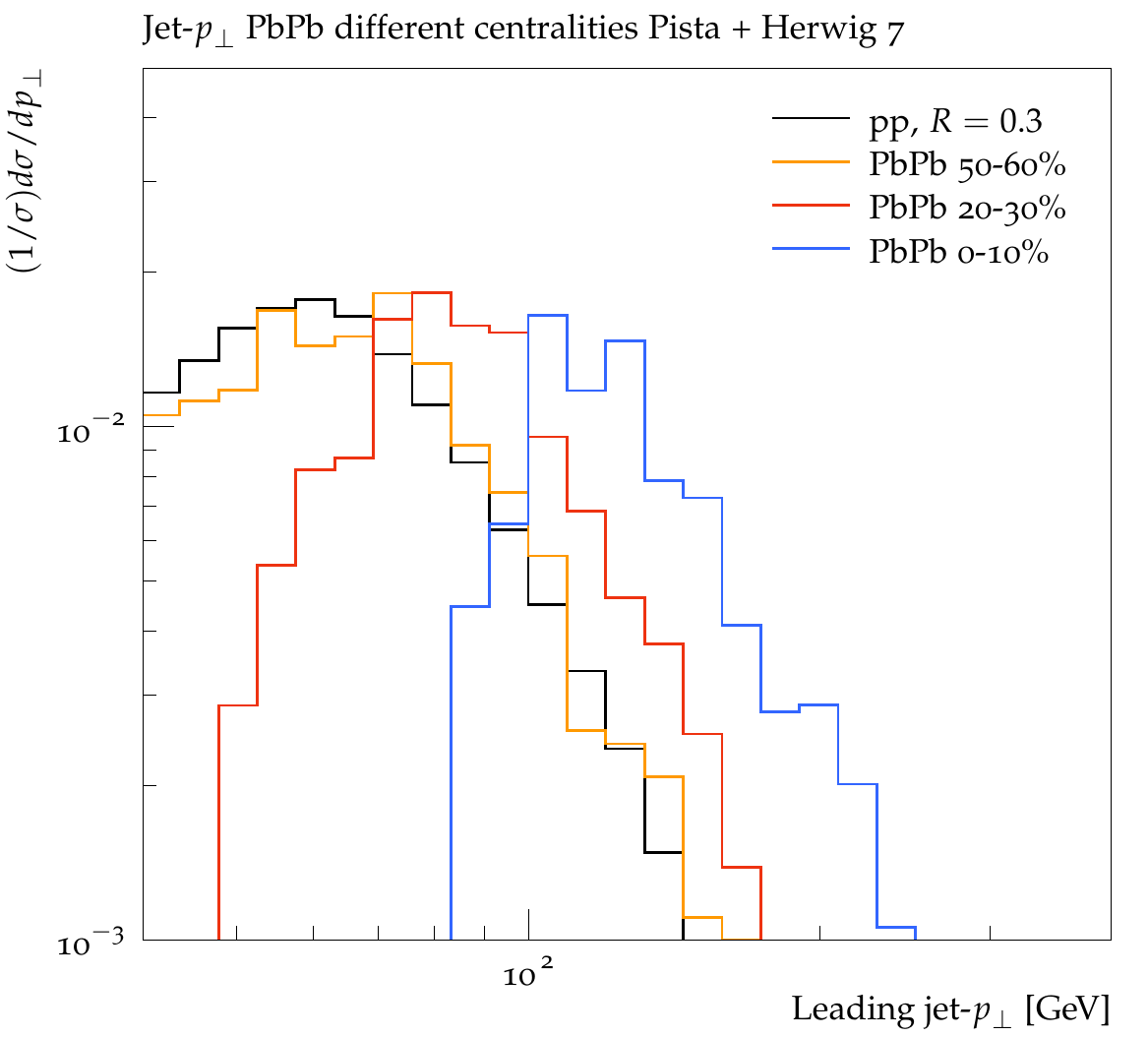}
	\end{center}
	\caption{\label{fig:jetpt-cent}Centrality binned leading jet-$p_\perp$ embedded in a {\it PbPb} background
	for Pista + Pythia 8 (left) and Herwig 7 (right) respectively.}
\end{figure}

In figure \ref{fig:jetpt-cent} we show the centrality binned leading jet-$p_\perp$ for Pista + Pythia~8(left)
and Pista + Herwig~7 (right). We note that the centrality definition in this figure is similar to what is
done in an experiment, \textit{i.e.}~by binning in the same centrality observable as used in section
\ref{sec:minbias}. This is a natural feature of the embedding procedure, as the $Z+jet$ event is embedded
in a full underlying event, but to the authors' knowledge not possible in any state-of-the-art jet quenching
simulation\footnote{The same feature is, by construction, of course available in the Angantyr addition
to Pythia 8~\cite{Bierlich:2018xfw}.}.

The centrality binned jet-$p_\perp$, has the striking feature that the spectrum is shifted upwards with
increasing centrality. This is again not a surprising feature, as the increased underlying event activity
will naturally manifest as "misidentified" leading jets. This is, however, an effect which must be handled
by experiments, and this demonstrates that this approach can be used directly to estimate the efficiency of
subtraction techniques as employed by experiments (in \textit{eg.} ref.~\cite{Sirunyan:2017jic}).

\subsection{Embedding with JEWEL}
Finally we demonstrate the ability to embed a signal event from a special purpose generator, with an 
underlying event from general purpose generators. For this purpose we use the JEWEL event generator 
\cite{Zapp:2013vla,KunnawalkamElayavalli:2016ttl}, which includes modifications to the jet from a 
Quark--Gluon Plasma. 

The first thing to note is the centrality definition. In JEWEL, forward event activity in heavy ion
collisions is not reproduced, and for the purpose of the medium calculation, a centrality (in percent) 
must be set. When embedded into a generated underlying event, the underlying event has a centrality 
based on event activity. These two centralities does not coincide. For the purpose of demonstration,
we therefore select a JEWEL centrality between 0-10\%, and show only events where this coincides with 
the underlying event centrality. For the purpose of a real physics analysis, the user ought to select
a finer centrality binning.

\begin{figure}
	\begin{center}
		\includegraphics[width=0.45\textwidth]{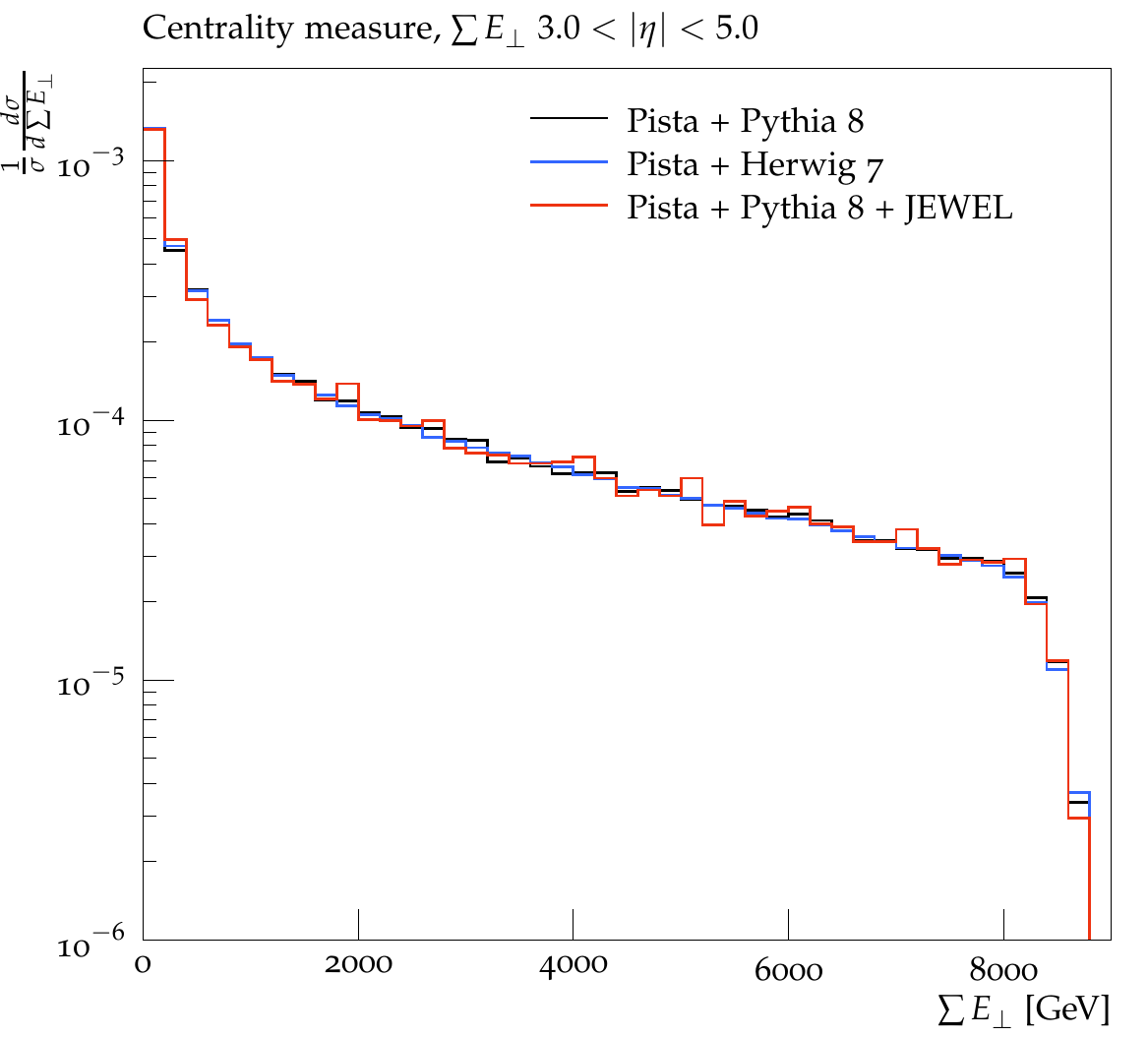}
		\includegraphics[width=0.45\textwidth]{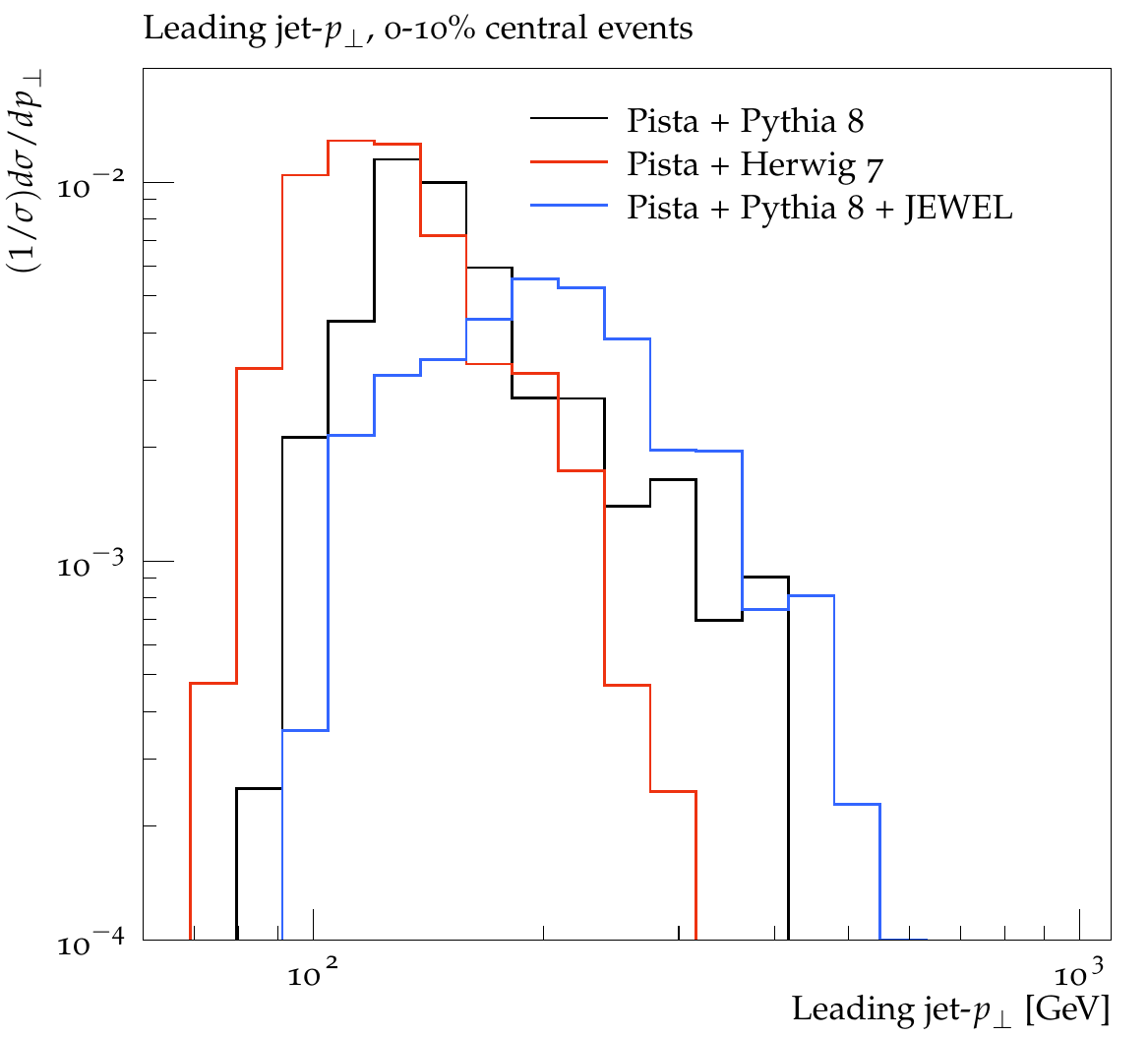}
	\end{center}
	\caption{\label{fig:centrality-hist}The centrality observable (left) used for PbPb collisions for
	Pista + Pythia 8, Herwig and (Pythia 8 + JEWEL) respectively. Transverse momentum (right) of the leading
	jet for 0-10\% central events using Pista + Pythia 8, Herwig and (Pythia 8 + JEWEL) respectively.}
\end{figure}

In figure \ref{fig:centrality-hist} (left), we show the PbPb centrality measure (as used in the previous sections)
for Pista + Pythia 8 and Herwig 7 respectively, as well as for JEWEL events embedded with Pista + Pythia 8. The generated
centrality measures are seen to coincide up to statistical errors. It is thus possible to select central JEWEL events
with underlying event activity matching the input centrality. This is done in figure \ref{fig:centrality-hist} (right),
where $p_\perp$ of the leading jet, in the same $Z+$ jet setup as in the previous section, is shown. The
JEWEL calculation is compared to the corresponding embedded vacuum calculations using Pythia 8 and Herwig 7.

\section{Conclusions} \label{sec:conclusion}

We have presented a generator independent framework for stacking {\it{pp}} collisions to obtain {\it{pA}} and {\it{AA}} collisions,
based on Angantyr.
Full, hadronized events are stacked, using the HepMC event record, allowing this framework to be very useful
for studies of jet quenching effects, as one can merge a quenched jet--event with an underlying event generated
with standard {\it{pp}} tools. 
While this is the first time the simulation of HI simulations containing Herwig 7 events are directly 
compared to Pythia 8, we plan to follow up the publication of the framework with physics studies 
of standard jet--quenching observables embedded in such an underlying event.

The framework is fully implemented in Python, and relies only on desired input event generators. For convenience,
we ship the framework with a set of already implemented Rivet analyses, useful for comparison to data.

\section*{Acknowledgements}
We thank Andy Buckley and Lukas Heinrich for providing open source Python wrappers to HepMC~\cite{pyhepmc,hepmcanalysis}.
This work was funded in part by the Swedish Research Council, contracts number 2012-02283 and 2017-0034, 
in part by the European Research Council (ERC) under the European Union’s Horizon 2020 research and 
innovation programme, grant agreement No 668679, and in part by the MCnetITN3 H2020 Marie Curie Initial
Training Network, contract 722104.

\appendix

\section{Correlated Collisions}
\label{app:correlatedCollisions}
In collisions with more than one nucleus on one of the beam sides  
it is possible that one of the participating nuclei is part in more than one sub-collisions.
A sub-collisions is in this section treated as an index pair labelling the participants of the 
colliding particles. 
We assume that we constructed an algorithm that allows us to identify a sub-collision
as hard  or additional and construct corresponding index pair vectors $\mathcal{H}$ and $\mathcal{A}$. 
Each index can at most appear once in $\mathcal{H}$ but multiple times in $\mathcal{A}$.
For a given index-pair $(i,j)$ in the $\mathcal{H}$ we  construct a method to tell 
the number $N$ of events to choose the hardest. 

\begin{enumerate}
\item If $\mathcal{A}$ is empty or does not contain $i$ in the left indices or $j$ in the right
indices we choose from N=1. 
\item Project $\mathcal{A}$ to a subset $\mathcal{A}_{(i||j)}$ where either $i$ is the left or
	$j$ is the right index. Model1: choose from the length of $\mathcal{A}_{(i||j)}$ as,
		$N=len(\mathcal{A}_{(i||j)} )+ 1$
\item For each pair $(k,l)$ in $\mathcal{A}_{(i||j)}$ add one to $N$ if the index ($k$ if $l==j$ else $l$)
	does not appear in the corresponding side of $\mathcal{H}$ and $1/2$ if it does. Model2: Choose from resulting $N+1$. 
\end{enumerate}

The resulting two models are very simple and also the appearance of the same indices in the additional
events needs to be studied but the purpose of this paper is to set the infrastructure and define the framework.

%% References
%% Usage of~\cite is as follows:
%%  ~\cite{key}         ==>>  [#]
%%  ~\cite[chap. 2]{key} ==>> [#, chap. 2]
%%


\bibliographystyle{elsarticle-num}
\bibliography{pista}






% \end{thebibliography}

\end{document}